\documentclass{birkmult}

\usepackage{amsmath,amssymb}
\usepackage{cite}

\newtheorem{thm}{Theorem}[section]
\newtheorem{lem}[thm]{Lemma}

\begin{document}

\title[Information spreading measures of orthogonal polynomials]{Information-theoretic-based spreading measures of orthogonal polynomials} %[[titulo corto]]

\author{J.S. Dehesa}
\address{Department of Atomic, Molecular and Nuclear Physics, University of Granada, Granada, Spain}
\email{dehesa@ugr.es}

\author{A. Guerrero}
\address{Department of Atomic, Molecular and Nuclear Physics, University of Granada, Granada, Spain}
\email{agmartinez@ugr.es}

\author{P. S\'anchez-Moreno}
\address{Department of Applied Mathematics, University of Granada, Granada, Spain}
\email{pablos@ugr.es}



\date{\today}

\begin{abstract}
The  macroscopic properties of  a quantum system strongly depend on the spreading of the physical eigenfunctions (wavefunctions) of its Hamiltonian operador over its confined domain. The wavefunctions are often controlled by classical or hypergeometric-type orthogonal polynomials (Hermite, Laguerre and Jacobi).  Here we discuss the spreading of these polynomials over its orthogonality interval by means of various information-theoretic quantities which grasp some facets of the polynomial distribution not yet analyzed. We consider the information-theoretic lengths closely related to the Fisher information and R\'enyi and Shannon entropies, which quantify the polynomial spreading far beyond the celebrated standard deviation.
\end{abstract}

\maketitle

\section{Introduction}

The special functions of applied mathematics and mathematical physics \cite{nikiforov_88,temme_96,olver_10} tower above the jungle of mathematical functions for its numerous, simple and elegant algebraic properties, what has facilitated their analytical manipulation and  application to solve and interpret a great deal of scientific and technological problems. Nevertheless there are still many open problems related to them whose solution would be very useful and interesting from both fundamental and applied standpoints. This is particularly the case for the information-theoretic properties, whose knowledge for the classical orthogonal polynomials in a real continuous variable has been reviewed in \cite{dehesa_jcam01} up to 2001 and their asymptotics in \cite{aptekarev_jcam10} up to 2010. The information theory of the discrete orthogonal polynomials and the special functions other than orthogonal polynomials is still in its infancy despite a few efforts on Airy and Bessel functions \cite{sanchezmoreno_jpa05,dehesa_jmp03,dehesa_ijbc02}, hyperspherical harmonics \cite{yanez_jmp99,dehesa_jmp07}, hypergeometric functions \cite{yanez_jmp08} and on orthogonal polynomials in a discrete variable \cite{aptekarev_ca09,dehesa_jdea11}.

According to the Hohenberg-Kohn density functional theory \cite{parr_89},
the physical and chemical properties of a quantum system are controlled by the distribution of the single-particle density of the system over its confined domain. This quantum-mechanical density is given by the Rakhmanov probability density of the special function which controls the single-particle wavefunction. Consequently, the physics and chemistry of the quantum systems essentially depend on the spreading of the special functions of applied mathematics over its domain of definition, which is usually fixed by the confining limits of the system. To identify and compute the most appropriate measures of the various facets of this spreading far beyond the celebrated root-mean-square or standard deviation is a main goal of the emerging theory of information of the special functions.

In this paper we are going to discuss the class of direct spreading measures (the standard deviation and the information-theoretic lengths of Fisher, R\'enyi and Shannon types) of the Rakhmanov probability density \cite{rakhmanov_mus77} of the orthogonal polynomials and to compute them for the Hermite, Laguerre and Jacobi polynomials. They have been shown to best quantify different aspects of the spreading of these polynomials all over its orthogonality interval. It is well known that these classical orthogonal polynomials controls the wavefunctions of numerous quantum systems \cite{nikiforov_88,bagrov_90,cooper_01,galindo_pascual_90,garciamartinez_pla09}, including the hydrogenic and oscillator-like systems, and they are very frequently used in modeling a great deal of scientific and technological phenomena. 

The structure of this review paper is the following. In Section \ref{sec_direct_spreading} we define the direct spreading measures of the orthogonal polynomials recently introduced \cite{hall_pra00,hall_pra01,hall_pra99,dehesa_jcam06,sanchezmoreno_jcam10,sanchezmoreno_jcam11,guerrero_jpa10} and we fix the notations which will be used in the rest of the paper. In Sections \ref{sec_standard_deviation}, \ref{sec_renyis_lengths} and \ref{sec_shannons_lengths} we give the values of the standard deviation and the Fisher length, and the R\'enyi and Shannon information-theoretic lengths, respectively, as well as the corresponding methodology to calculate them. Then, in Section \ref{sec_numerical_discussion} the numerical analysis of these quantities is done for some particular polynomials. Finally, some conclusions are given and various open problems are posed.

\section{Direct spreading measures of orthogonal polynomials}
\label{sec_direct_spreading}

Let us here describe the direct spreading measures of the real hypergeometric-type orthonormal polynomials, i.e. the polynomials which satisfy the orthogonality relation
\begin{equation}
\int_{\Delta} p_n(x) p_m(x) \omega(x) dx =\delta_{n,m},\quad m,n\in\mathbb{N},
\label{eq_orthogonality_relation}
\end{equation}
where the weight function $\omega(x)$ has the expressions
\begin{align}
\omega_H(x)&=e^{-x^2}, \nonumber\\
\omega_L(x)&=x^\alpha e^{-x}, \quad (\alpha>-1) \label{eq_weight_functions}\\
\omega_J(x)&=(1-x)^\alpha (1+x)^\beta, \quad (\alpha,\beta>-1) \nonumber
\end{align}
for the three canonical families of Hermite $H_n(x)$, Laguerre $L_n^{(\alpha)}(x)$ and Jacobi $P_n^{(\alpha,\beta)}(x)$, respectively.  The spreading measures which quantify the different facets of the distribution of these polynomials all over the orthogonality interval are given by the corresponding measures of their associated Rakhmanov probability density, which is defined by 
\begin{equation}
\rho_n(x) =p_n^2(x)\omega(x),
\label{eq_rakhmanov_density}
\end{equation}
as defined by this mathematician \cite{rakhmanov_mus77}, who first discovered that this density governs the asymptotic ($n\to+\infty$) behavior of the ratio of two polynomials with consecutive orders. It is worth saying here that this normalized-to-unity probability density characterizes the stationary states of a large family of quantum-mechanical potentials \cite{nikiforov_88,galindo_pascual_90,bagrov_90,cooper_01,garciamartinez_pla09}.

The class of direct spreading measures of a probability density \cite{hall_pra00,hall_pra01,hall_pra99} include the standard deviation and the information-theoretic lengths of Fisher, R\'enyi and Shannon types.
The standard deviation  of the polynomial $p_n(x)$ is given by the following root-mean-square
\begin{equation}
(\Delta x)_n =\left(\langle x^2\rangle_n-\langle x\rangle^2_n\right)^\frac12,
\label{eq_standard_deviation}
\end{equation}
where the expectation value of a function $f(x)$ is defined by  
\begin{equation}
\langle f(x)\rangle_n=\int_\Delta f(x) \rho_n(x) dx.
\label{eq_expectation_value_f}
\end{equation}

The R\'enyi information-theoretic lengths \cite{hall_pra99} of the (Rakhmanov density associated to) the polynomial $p_n(x)$ is defined as 
\begin{equation}
\mathcal{L}_q^R[\rho_n]=\left\langle \left[\rho_n(x)\right]^{q-1}\right\rangle^{-\frac{1}{q-1}}
=\left\{\int_\Delta \left[\rho_n(x)\right]^q dx\right\}^{-\frac{1}{q-1}},\quad q>0,q\ne 1,
\label{eq_rengyi_length}
\end{equation}
where $R_q[\rho_n]$ denotes the $q$th-order R\'enyi entropy \cite{renyi_1}.  Let us highlight the case $q=2$, which corresponds to the Onicescu-Heller length (closely related to the notions of disequilibrium, inverse participation ratio, Br\"ukner-Zeilinger entropy and quantum entanglement in various contexts) given by
\begin{equation}
\mathcal{L}_2^R[\rho_n]=\left\langle \left[\rho_n(x)\right]\right\rangle^{-1}
=\left\{\int_\Delta \left[\rho_n(x)\right]^2 dx\right\}^{-1},
\label{eq_heller_length}
\end{equation}
and the Shannon length \cite{hall_pra99}, which corresponds to the limiting case $q\to 1$:
\begin{equation}
N[\rho_n]=\lim_{q\to 1} \mathcal{L}_q^R[\rho_n]=\exp(S[\rho_n])
=\exp\left\{-\int_\Delta \rho_n(x) \ln\rho_n(x) dx\right\},
\label{eq_shannon_length}
\end{equation}
where $S[\rho_n]$ denotes the Shannon information entropy \cite{shannon_1}. 

  Now let us consider a direct spreading measure which is qualitatively different from the previous ones, the Fisher length \cite{hall_pra00,hall_pra01}; it is defined by
\begin{equation}
(\delta x)_n\equiv\frac{1}{\sqrt{F[\rho_n]}}\equiv \left\langle \left[ \frac{d}{dx}\ln\rho_n(x)\right]^2\right\rangle^{-\frac{1}{2}}=\left\{\int_{\Delta}dx\frac{[\rho_n'(x)]^2}{\rho_n(x)}\right\}^{-\frac{1}{2}},
\label{eq_fisher_length}
\end{equation}
where $F[\rho_n]$ denotes the Fisher information \cite{frieden_1} of the polynomial $p_n$. It is interesting to remark that this quantity is a functional of the derivative of the Rakhmanov density of the polynomial. So, the Fisher length is very sensitive to the polynomial oscillations. It is a local quantity in the sense that it measures the pointwise concentration of the probability over the orthogonality interval, and it quantifies the gradient content of the Rakhmanov density providing (i) a quantitative estimation of the oscillatory character of the density and the polynomials, and (ii) the bias to particular points of the interval, so that it measures the degree of local disorder.

In contrast, the standard deviation and the R\'enyi and Shannon lengths are global spreading measures because they are power-like (standard deviation and R\'enyi lengths) and logarithmic (Shannon length) functionals of the density. Moreover, the standard deviation is a measure of separation of the probability cloud from a particular point of the support interval (namely, the mean value or centroid) while the R\'enyi and Shannon lengths are measures of the extent to which the density (so, the polynomial) is in fact concentrated. 

Finally, let us mention that all the four direct spreding measures share the following properties \cite{hall_pra00,hall_pra01,hall_pra99}: same units as the random variable, translation and reflection invariance and linear scaling under adequate boundary conditions, and vanishing when the density tends to a delta function. Moreover they fulfil an uncertainty property \cite{kennard_zp27,bialynicki_cmp75,zozor_jpa08,sanchezmoreno_jpa11} and the Cram\'er-Rao \cite{hall_pra00} and Shannon \cite{shannon_1} inequalities given by
\begin{equation}
(\delta x)_n \le (\Delta x)_n, \quad\text{and}\quad N[\rho_n]\le (2\pi e)^\frac12 (\Delta x)_n,
\label{eq_cramer_rao_inequality}
\end{equation}
respectively.

\section{Standard deviation and Fisher's length of classical orthogonal polynomials}
\label{sec_standard_deviation}

In this Section we give the values of the standard deviation and the Fisher length of the Hermite, Laguerre and Jacobi polynomials in terms of their degree and the characterizing parameters, as well as the methodology used for their analytical calculation.

First let us show that the standard deviation $\Delta x$, given by (\ref{eq_standard_deviation}), has the value
\begin{equation}
(\Delta x)_n=\sqrt{n+\frac12},
\label{eq_standard_deviation_hermite}
\end{equation}
for the Hermite polynomials $H_n(x)$,
\begin{equation}
(\Delta x)_{n,\alpha}=\sqrt{2n^2+2(\alpha+1)n+\alpha+1},
\label{eq_standard_deviation_laguerre}
\end{equation}
for the Laguerre polynomials $L_n^{(\alpha)}(x)$, $\alpha>-1$, and
\begin{multline}
\left(\Delta x \right)_{n,\alpha,\beta}=
\left[\frac{4(n+1)(n+\alpha+1)(n+\beta+1)(n+\alpha+\beta+1)}{(2n+\alpha+\beta+1)(2n+\alpha+\beta+2)^2 (2n+\alpha+\beta+3)}\right.\\
+\left.\frac{4n(n+\alpha)(n+\beta)(n+\alpha+\beta)}{(2n+\alpha+\beta-1)(2n+\alpha+\beta)^2 (2n+\alpha+\beta+1)}\right]^{1/2},
\label{eq_standard_deviation_jacobi}
\end{multline}
for the Jacobi polynomials $P_n^{(\alpha,\beta)}(x)$, with $\alpha,\beta>-1$. These results can be obtained from Eqs. (\ref{eq_standard_deviation}) and (\ref{eq_expectation_value_f}) and the three-term recurrence relation of the polynomials \cite{dehesa_jcam06} (see also \cite{sanchezmoreno_jcam10}, \cite{sanchezmoreno_jcam11} and \cite{guerrero_jpa10} for an alternative proof of the Hermite, Laguerre and Jacobi cases, respectively).

On the other hand, the Fisher information-theoretic length $\delta x$, defined by (\ref{eq_fisher_length}), has the value
\begin{equation}
(\delta x)_n=\frac{1}{\sqrt{4n+2}},
\label{eq_fisher_length_hermite}
\end{equation}
for Hermite polynomials $H_n(x)$,
\begin{eqnarray}
(\delta x)_{n,\alpha}=
\left\{
\begin{array}{cc}
\frac{1}{\sqrt{4n+1}}; & \alpha=0,\\[2mm]
\sqrt{\frac{\alpha^2-1}{(2n+1)\alpha+1}};&\alpha>1,\\[2mm]
0;&  \alpha \in (-1,+1],\alpha\neq 0.\\
\end{array}
\right.
\label{eq_fisher_length_laguerre}
\end{eqnarray}
for Laguerre polynomials $L_n^{(\alpha)}(x)$, $\alpha>-1$, and
\begin{equation}
(\delta x)_{n,\alpha,\beta}=\frac{1}{\sqrt{F\left[\rho_{n,\alpha,\beta}\right]}},
\label{eq_fisher_length_jacobi}
\end{equation}
where the Fisher information $F\left[\rho_{n,\alpha,\beta}\right]$ has the expression 
\begin{eqnarray*}
F\left[\rho_{n,\alpha,\beta}\right]=
\left\{
\begin{array}{ll}
2n(n+1)(2n+1), & \alpha,\beta=0, \\
\frac{2n+\beta+1}{4}\left[\frac{n^2}{\beta+1}+n+(4n+1)(n+\beta+1)+\frac{(n+1)^2}{\beta-1}\right],
& \alpha=0, \beta>1, \\
\frac{2n+\alpha+\beta+1}{4(n+\alpha+\beta-1)}\left[n(n+\alpha+\beta-1)
\left(\frac{n+\alpha}{\beta+1}+2+\frac{n+\beta}{\alpha+1}\right)\right.& \\
+\left.(n+1)(n+\alpha+\beta)\left(\frac{n+\alpha}{\beta-1}+2+\frac{n+\beta}{\alpha-1}\right)\right],
&\alpha, \beta>1, \\
\infty, & {\rm otherwise,}
\end{array}
\right.
\end{eqnarray*}
for Jacobi polynomials $P_n^{(\alpha,\beta)}(x)$, with $\alpha,\beta>-1$. These expressions follow from Eq. (\ref{eq_fisher_length}) and the algebraic determination of the Fisher information of the classical orthogonal polynomials \cite{sanchezruiz_jcam05,yanez_jmp08}.

The comparison of Eqs. (\ref{eq_standard_deviation_hermite})-(\ref{eq_standard_deviation_jacobi}) and (\ref{eq_fisher_length_hermite})-(\ref{eq_fisher_length_jacobi}) allows us to point out that the Cram\'er-Rao inequality $\delta x\le \Delta x$, valid for all continuous probability densities on the whole real line, is indeed fulfilled not only in the Hermite case but also in the Laguerre and Jacobi cases. It is easy to verify that the Cram\'er-Rao inequality is satisfied for $\alpha=0$, and for $\alpha=\beta=0$ and $\alpha=0,\beta>0$, respectively. And the corresponding Rakhmanov densities for $\alpha>1$ and $\alpha,\beta>1$,  with supports $[0,+\infty)$ and $[-1,+1]$, respectively, vanish at $x=0$ and $x=\pm 1$, so that they can be considered as continuous on the whole real line. Moreover, the Cram\'er-Rao product $(\delta x)_n(\Delta x)_n=\frac12$ for Hermite polynomials. 
Finally, it is worth noting that the asymptotic ($n\to+\infty$) behaviour of this product is
\begin{eqnarray*}
(\delta x)_{n,\alpha} (\Delta x)_{n,\alpha}\simeq
\left\{
\begin{array}{cc}
\sqrt{\frac{n}{2}}; & \alpha=0,\\[2mm]
\sqrt{\frac{\alpha^2-1}{\alpha}n};&\alpha>1,\\[2mm]
0;&  \alpha \in (-1,+1],\alpha\neq 0.\\
\end{array}
\right.
%\label{eq_fisher_length_laguerre}
\end{eqnarray*}
for the Laguerre polynomials $L_n^{(\alpha)}(x)$, and
\begin{eqnarray*}
(\delta x)_{n,\alpha,\beta} (\Delta x)_{n,\alpha,\beta}\simeq
\left\{
\begin{array}{cc}
(2n)^{-\frac32}; & \alpha,\beta=0,\\[2mm]
\left(\frac{1}{\beta+1}+\frac{1}{\beta-1}+4\right)^{-\frac12} n^{-\frac32};&\alpha=0,\beta>1,\\[2mm]
\left(\frac{1}{\beta+1}+\frac{1}{\beta-1}+\frac{1}{\alpha+1}+\frac{1}{\alpha-1}\right)^{-\frac12} n^{-\frac32};&\alpha>1,\beta>1,\\[2mm]
0;& \text{otherwise},\\
\end{array}
\right.
%\label{eq_fisher_length_laguerre}
\end{eqnarray*}
for the Jacobi polynomials $P_n^{(\alpha,\beta)}(x)$.

\section{R\'enyi's lengths of classical orthogonal polynomials}
\label{sec_renyis_lengths}

In this Section we show the value of the R\'enyi lengths $\mathcal{L}_q^R[\rho_n]$, with half-integer order $q$, of Hermite, Laguerre and Jacobi polynomials obtained by a combinatorial methodology based on the use of the multivariate Bell polynomials \cite{comtet_74}, and the values of these quantities obtained for the Laguerre polynomials by means of an algebraic methodology based on the linearization relation of Srivastava-Niukkanen \cite{srivastava:mcm03} and the use of Lauricella functions \cite{srivastava_85}.

\subsection{Combinatorial approach}

This approach is based on the following lemma
\begin{lem}[Ref. \cite{sanchezmoreno_jcam10}]
The $p$-th power of the polynomial
\begin{equation}
y_n(x)=\sum_{t=0}^n c_t x^t
\label{eq_polynomial_explicit}
\end{equation}
is given by
\begin{equation}
[y_n(x)]^p =\sum_{t=0}^{np} \frac{p!}{(t+p)!} B_{t+p,p}(c_0,2!c_1,\ldots,(t+1)!c_t)x^t,
\end{equation}
with $c_i=0$ for $i>n$. The $B$-symbols denote the Bell polynomials of Combinatorics \cite{comtet_74} which are given by
\begin{equation}
B_{m,l}(c_1,c_2,\ldots,c_{m-l+1})=\sum_{\hat{\pi}(m,l)} \frac{m!}{j_1! j_2!\cdots j_{m-l+1}!}\left(\frac{c_1}{1!}\right)^{j_1} \left(\frac{c_2}{2!}\right)^{j_2}\cdots \left(\frac{c_{m-l+1}}{(m-l+1)!}\right)^{j_{m-l+1}}
\end{equation}
where the sum runs over all partitions $\hat{\pi}(m,l)$ such that
\begin{equation}
j_1+j_2+\cdots+j_{m-l+1}=l, \quad \text{and}\quad j_1+2j_2+\cdots+(m-l+1)j_{m-l+1}=m.
\end{equation}
\end{lem}

From the definition (\ref{eq_rengyi_length}) of the R\'enyi length of the Rakhmanov density (\ref{eq_rakhmanov_density}) of the classical orthogonal polynomials, the explicit expressions (\ref{eq_polynomial_explicit}) of these polynomials and this lemma, we obtain the following $\mathcal{L}_q^R[\rho_n]$ values for $2q\in\mathbb{N}, q>2$:
\begin{itemize}
\item[(i)] Hermite polynomials $H_n(x)$ \cite{sanchezmoreno_jcam10}
\begin{equation}
\mathcal{L}_q^R[\rho_n]=\left\{W_q[\rho_n]\right\}^{-\frac{1}{q-1}}=\left(\sum_{j=0}^{nq}\frac{\Gamma\left(j+\frac12\right)}{q^{j+\frac12}}\frac{(2q)!}{(2j+2q)!}
B_{2j+2q,2q}(c_0^{(n)},2!c_1^{(n)},\ldots,(2j+1)!c_{2j}^{(n)})\right)^{-\frac{1}{q-1}},
\label{eq_renyi_length_hermite_bell}
\end{equation}

where the expansion coefficients $c_t$ are given by
\begin{equation}
c_t^{(n)}=\frac{(-1)^{\frac{3n-t}{2}}n!}{\left(2^n n!\sqrt{\pi}\right)^\frac12}\frac{2^t}{\left(\frac{n-t}{2}\right)!t!}\frac{(-1)^t+(-1)^n}{2},
\end{equation}

\item[(ii)] Laguerre polynomials $L_n^{(\alpha)}(x)$ \cite{sanchezmoreno_jcam11}
\begin{equation}
\mathcal{L}_q^R\left[\rho_{n,\alpha}\right]= \left[ \sum_{k=0}^{2nq}\frac{\Gamma(\alpha q+k+1)}
{q^{\alpha q+k+1}} \frac{(2q)!}{(k+2q)!}
B_{k+2q,2q}\left(c_0^{(n,\alpha)},2!c_1^{(n,\alpha)},...,(k+1)!c_k^{(n,\alpha)}\right)\right]^{-\frac{1}{q-1}},
\label{eq_renyi_length_laguerre_bell}
\end{equation}
with the Laguerre expansion coefficients
\begin{equation}
c_t^{(n,\alpha)}=\sqrt{\frac{\Gamma(n+\alpha+1)}{n!}}\frac{(-1)^t}{\Gamma(\alpha+t+1)}\binom{n}{t},
\end{equation}

\item[(iii)] Jacobi polynomials $P_n^{(\alpha,\beta)}(x)$ \cite{guerrero_jpa10}
\begin{equation}
\mathcal{L}_q^R\left[\rho_{n,\alpha,\beta}\right]=\left(\sum_{k=0}^{2nq}\frac{(2q)!}{(k+2q)!}
B_{k+2q,2q}\left(c_0^{(n,\alpha,\beta)},2!c_1^{(n,\alpha,\beta)},\ldots,(k+1)!c_k^{(n,\alpha,\beta)}\right)
\mathcal{I}(k,q,\alpha,\beta)
\right)^{-\frac{1}{q-1}},
\label{eq_renyi_length_jacobi_bell}
\end{equation}
where
\[
\mathcal{I}(k,q,\alpha,\beta) =
\frac{(-1)^k 2^{1+\alpha q+\beta q}\Gamma(\alpha q+1)\Gamma(\beta q+1)}{\Gamma(\alpha q+\beta q+2)}
\,_2F_1 \left(
\begin{array}{l}
 -k,1+\beta q\\
2+(\alpha+\beta)q
\end{array}
;2\right).
\]
and with the Jacobi expansion coefficients
\begin{multline}
c_{t}^{(n,\alpha,\beta)}=\sqrt{\frac{\Gamma(\alpha +n+1)(2n+\alpha +\beta +1)}{n!2(\alpha +\beta +1)\Gamma(\alpha +\beta +n+1)\Gamma(n+\beta +1)}}\\
\times
\sum_{i=t}^{n}(-1)^{i-t}\binom{n}{i}\binom{i}{t}
\frac{\Gamma(\alpha +\beta +n+i+1)}{2^{i}\Gamma(\alpha +i+1)}.
\end{multline}

\end{itemize}

From relations (\ref{eq_renyi_length_hermite_bell})-(\ref{eq_renyi_length_jacobi_bell}) we can obtain all the R\'enyi lengths $\mathcal{L}_q^R[\rho]$ with $2q\in \mathbb{N}$ of the three real classical orthogonal polynomials. In particular, for $q=2$ we obtain the following values for the Onicescu-Heller information-theoretic length $\mathcal{L}_2^R[\rho_n]$ (see Eq. (\ref{eq_heller_length})):
\[
\mathcal{L}_2^R[\rho_0]=\sqrt{2\pi},\quad \mathcal{L}_2^R[\rho_1]=\frac43\sqrt{2\pi},\quad \mathcal{L}_2^R[\rho_2]=\frac{64}{41}\sqrt{2 \pi}
\]
of the first few Hermite polynomials with degrees $n=0,1,2$,
\begin{equation}
\mathcal{L}_2^R[\rho_{0,\alpha}]=\left(\frac{2^{2\alpha+1} \left(\Gamma(\alpha+1)\right)^2}{\Gamma(2\alpha+1)}\right)^\frac12,
\label{eq_heller_length_laguerre_n0_bell}
\end{equation}
\begin{equation}
\mathcal{L}_2^R[\rho_{1,\alpha}]=\left(\frac{2^{2\alpha+3}\left(\Gamma(\alpha+2)\right)^2}{(1+\alpha)(2+3\alpha)\Gamma(2\alpha+1)}\right)^\frac12
\label{eq_heller_length_laguerre_n1_bell}
\end{equation}
of the Laguerre polynomials $L_n^{(\alpha)}(x)$ with $n=0,1$, and
\begin{equation}
\mathcal{L}_2^R[\rho_{0,\alpha,\beta}]=\left[
\left(\frac{\Gamma(\alpha+\beta+2)}{2^{\alpha+\beta+1}\Gamma(\alpha+1)\Gamma(\beta+1)}\right)^q
\frac{2^{1+2\alpha+2\beta}\Gamma(2\alpha+1)\Gamma(2\beta+1)}{\Gamma(2\alpha+2\beta+2)}
\right]^{-1},
\end{equation}
\begin{equation}
\mathcal{L}_2^R[\rho_{1,\alpha,\beta}]=\left[
\sum_{k=0}^{4} \binom{4}{n} c_0^{4-k}c_1^k \mathcal{I}(k,2,\alpha,\beta)
\right]^{-1},
\end{equation}
of the Jacobi polynomials $P_n^{(\alpha,\beta)}(x)$ with $n=0$ and $1$.

\subsection{Algebraic approach}

This approach is based on the linearization relation of an arbitrary product of classical orthogonal polynomials of the same type which would allow us to express the $q$th-power of the polynomials in terms of the polynomials themselves, so that the power functional involved in the R\'enyi lengths (\ref{eq_rengyi_length}) of the associated Rakhmanov density would be solvable by using the orthogonality relation. Such a linearization relation is known only for Laguerre polynomials \cite{srivastava:mcm03,sanchezmoreno_jcam11}, to the best of our knowledge. Indeed, the linearization relations found by the Srivastava-Niukkanen for arbitrary products of Laguerre polynomials \cite{srivastava:mcm03} have allowed us to find \cite{sanchezmoreno_jcam11} the expansion relation
\begin{equation}
(qt)^{2q} \left[ L_n^{(\alpha)}(t)\right]^{2q}=\sum_{k=0}^{\infty}
\Theta_k \left(\alpha q,0,2q,\{n\},\{\alpha\},\left\{\frac{1}{q}\right\}\right) L_k^{(0)}(qt),
\label{eq_qt2q_Ln2q_expansion}
\end{equation}
with the $\Theta_k$-coefficients given by
\begin{multline}
\Theta_k \left(\alpha q,0,2q,\{n\},\{\alpha\},\left\{\frac{1}{q}\right\}\right) = \Gamma(\alpha q+1)\binom{n+\alpha}{n}^{2q}\\ \times F_A^{(2q+1)}\left(\alpha q+1;-n,\cdots,-n;-k;
\alpha+1\cdots,\alpha+1,1;\frac{1}{q},\cdots,\frac{1}{q},1\right).
\label{eq_theta_k}
\end{multline}
where $F_A^{(r+1)}$ denotes a Lauricella's hypergeometric function of $r+1$ variables \cite{srivastava_85}. The combination of Eqs. (\ref{eq_rakhmanov_density}), (\ref{eq_rengyi_length}), (\ref{eq_qt2q_Ln2q_expansion}) and (\ref{eq_theta_k}) yields the following value for the $q$th-order R\'enyi length of the Laguerre polynomial
\begin{multline}
\mathcal{L}_q^R \left[\rho_{n,\alpha}\right]=
\left[ \left(\frac{n!}{\Gamma(\alpha+n+1)}\right)^q
 \frac{1}{q^{\alpha q+1}} \Gamma (\alpha q+1) \binom{n+\alpha}{n}^{2q}\right.\\
\times F_A^{(2 q+1)} 
\left.
\left(
\alpha q+1,-n,\cdots,-n,0;\alpha+1,...,\alpha+1,1; \frac{1}{q},\cdots,\frac{1}{q},1\right)\right]^{-\frac{1}{q-1}},
\end{multline}
for every $q>0$ such that $2q\in\mathbb{N}$. As particular cases, this expression gives the value
\begin{equation}
\mathcal{L}_q^R[\rho_{0,\alpha}]=\left[\frac{1}{\Gamma(\alpha+1)^q}
\frac{\Gamma(\alpha q+1)}{q^{\alpha q+1}}\right]^{-\frac{1}{q-1}},
\label{eq_renyi_length_laguerre_n0_lauricella}
\end{equation}
and
\begin{equation}
\mathcal{L}_q^R\left[\rho_{1,\alpha}\right]=\left[\frac{\Gamma(\alpha q+1)(1+\alpha)^{2q}}{(\Gamma(\alpha+2))^q q^{\alpha q+1}}
\,_2F_0\left(
\begin{array}{c}
-2q,\alpha q+1\\
-
\end{array}
;\frac{1}{q(\alpha+1)}
\right)
\right]^{-\frac{1}{q-1}}.
\label{eq_renyi_length_laguerre_n1_lauricella}
\end{equation}
for the Laguerre polynomials $L_n^{(\alpha)}(x)$ with lowest degrees $n=0,1$. It is worth remarking that Eqs. (\ref{eq_renyi_length_laguerre_n0_lauricella}) and (\ref{eq_renyi_length_laguerre_n1_lauricella}) with $q=2$ boil down to the previous values (\ref{eq_heller_length_laguerre_n0_bell}) and (\ref{eq_heller_length_laguerre_n1_bell}), respectively, of the Onicescu-Heller lengths.

\section{Shannon's length of classical orthogonal polynomials}
\label{sec_shannons_lengths}

The Shannon length (\ref{eq_shannon_length}) of the Rakhmanov density (\ref{eq_rakhmanov_density}) of the classical orthogonal polynomials of the classical orthogonal polynomials $N[\rho_n]=\exp(S[\rho_n])$ have not yet been analytically calculated in terms of the degree $n$ and the characterizing parameters, mainly because it is a logarithmic functional of the polynomials. Here, we show ``only'' its asymptotic $(n\to +\infty)$ behaviour and the common linear asymptotical relation with the standard deviation. Moreover, we also give sharp bounds to $N[\rho_n]$ in terms of various expectation values $\langle f(x)\rangle$.

\subsection{Asymptotics}

In a series of works initiated in 1994 and recently summarized in \cite{aptekarev_jcam10}, Aptekarev et al have studied the asymptotic behaviour of the Shannon entropy
\[
S[\rho_n]=-\int \omega(x)\rho_n(x)\log\rho_n(x) dx
\]
of the Rakhmanov density $\rho_n(x)$ defined by Eq. (\ref{eq_rakhmanov_density}) for the classical orthogonal polynomials given by Eqs. (\ref{eq_orthogonality_relation}) and (\ref{eq_weight_functions}). They have found the values
\[
S[\rho_n]=\log\sqrt{2n}+\log\pi-1+o(1),
\]
\[
S[\rho_{n,\alpha}]=(\alpha+1)\log n-\alpha \psi(\alpha+n+1)-1+\log(2\pi)+o(1),
\]
and
\begin{equation*}
S[\rho_{n,\alpha,\beta}]=\log\pi-1+o(1)
\end{equation*}
for the Hermite, Laguerre and Jacobi polynomials, respectively, by use of powerful, highbrow concepts and techniques of approximation theory \cite{aptekarev_rassm95}. Taking these results into (\ref{eq_shannon_length}), we obtain the following asymptotical values
\begin{equation}
N[\rho_n]\simeq\frac{\pi\sqrt{2n}}{e}
\label{eq_shannon_length_hermite_asymptotics}
\end{equation}
\begin{equation}
N[\rho_{n,\alpha}]\simeq \frac{2\pi n}{e}
\label{eq_shannon_length_laguerre_asymptotics}
\end{equation}
and
\begin{equation}
N[\rho_{n,\alpha,\beta}]\simeq \frac{\pi}{e}
\label{eq_shannon_length_jacobi_asymptotics}
\end{equation}
for the Shannon length of Hermite, Laguerre and Jacobi polynomials, respectively.

The comparison of Eqs. (\ref{eq_shannon_length_hermite_asymptotics})-(\ref{eq_shannon_length_jacobi_asymptotics}) with the asymptotical values of the standard deviation $\Delta x$ given by Eqs. (\ref{eq_standard_deviation_hermite})-(\ref{eq_standard_deviation_jacobi}) for Hermite, Laguerre and Jacobi polynomials, respectively, yields the following linear relation
\begin{equation}
N[\rho]\simeq \frac{\pi\sqrt{2}}{e}\Delta x\simeq 1.6389 \Delta x,
\label{eq:missing_one}
\end{equation}
which is common for all three cases \cite{devicente_04} (see also \cite{sanchezmoreno_jcam10,sanchezmoreno_jcam11,guerrero_jpa10}). It is interesting to mention here that this relation also holds for the whole class of Bernstein-Szeg\"o polynomials \cite{devicente_04} but it is not true for arbitrary orthogonal polynomials, since e.g. it is violated for Freud polynomials. Indeed, the Shannon length and the standard deviation has a quadratic relation in the Freud case \cite{devicente_04}.

\subsection{Upper bounds}

Since the exact value of the Shannon length $N[\rho_n]$ of the classical orthogonal polynomials cannot be calculated in terms of the degree $n$ and the characterizing parameters, it is natural to look for analytical upper bounds to this quantity as simple and accurate as possible. This has been done in 2010 \cite{sanchezmoreno_jcam10,sanchezmoreno_jcam11,guerrero_jpa10} in terms of expectation values $\langle x^k\rangle_n$ of the associated Rakhmanov density given by (\ref{eq_rakhmanov_density}).

In the Hermite and Laguerre cases \cite{sanchezmoreno_jcam10,sanchezmoreno_jcam11} we have used an optimized information-theoretic technique based on the non-negativity of the Kullback-Leibler functional of the Rakhmanov density of the polynomial and a probability density of exponentially decreasing type of the form $\exp\left(-x^k)\right)$. It is found that
\begin{equation*}
N[\rho_n]\le \frac{2(ek)^\frac{1}{k}}{k}\Gamma\left(\frac{1}{k}\right) \langle x^k\rangle_n^\frac{1}{k}; \quad k=2,4,\ldots
\end{equation*}
and
\begin{equation*}
N[\rho_{n,\alpha}]\le \frac{\Gamma\left(\frac{1}{b}\right)(be)^\frac{1}{b}}{b}\langle x^b\rangle^\frac{1}{b};\quad b>0,
\end{equation*}
for the R\'enyi length of Hermite and Laguerre polynomials, respectively. For completeness we give here the values
\begin{equation}
\label{eq_1}
\langle x^k\rangle_n=\left\{
\begin{array}{ll}
\frac{k!}{2^k \Gamma\left(\frac{k}{2}+1\right)}\,_2F_1\left(\left.
\begin{array}{c}
-n,-\frac{k}{2}\\
1
\end{array}
\right| 2
\right),
& \text{even } k\\
0, & \text{odd }k
\end{array}
\right.,
\end{equation}
and
\begin{equation}
\label{eq_2}
\langle x^k\rangle_{n,\alpha}=\frac{n!\Gamma(k+\alpha+1)}{\Gamma(n+\alpha+1)} \sum_{r=0}^n
\binom{k}{n-r}^2\binom{k+\alpha+r}{r},
\end{equation}
of the ordinary moments of Hermite and Laguerre polynomials, respectively. See \cite{sanchezmoreno_jcam10} and \cite{sanchezmoreno_jcam11} for further discussion about these bounds. Let us point out here, for completeness, that the results (\ref{eq_standard_deviation_hermite}) and
(\ref{eq_standard_deviation_laguerre}) may be also obtained from Eq. (\ref{eq_standard_deviation}) together with Eqs.        
 (\ref{eq_1}) and (\ref{eq_2})
with $k=1$ and $2$, respectively.

In the Jacobi case, beyond the general upper bound $N[\rho]\le 2$ valid for any probability density $\rho(x)$ with the support interval $[-1,+1]$, we \cite{guerrero_jpa10} have variationally found various upper bounds in terms of the following expectation values: $\langle x\rangle$, $\langle x^2\rangle$, $\langle \log x^2\rangle$, $\langle\log(1\pm x)\rangle$, $\langle\log(1-x^2)\rangle$. The analytical and numerical analysis of these bounds are discussed in detail in Ref. \cite{guerrero_jpa10}.

\section{Numerical discussion}
\label{sec_numerical_discussion}

\begin{figure}
\begin{center}
\includegraphics[width=10cm]{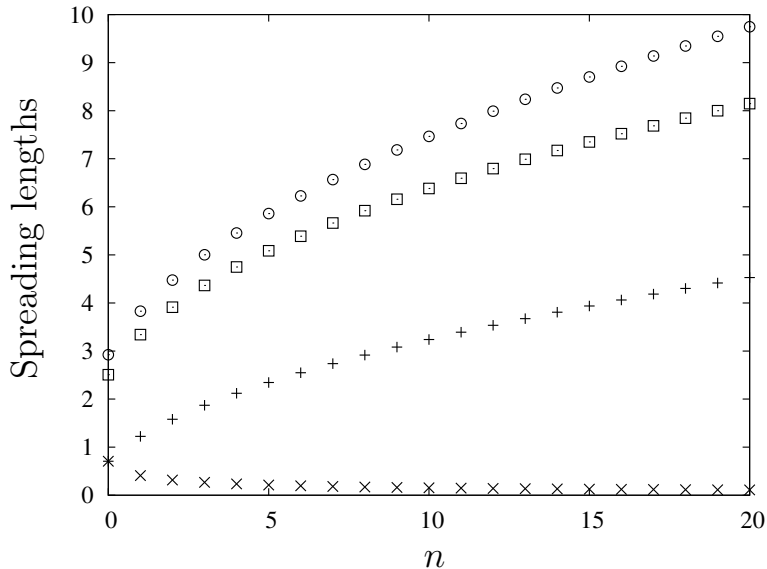}
\caption{Standard deviation $\Delta x$ ($+$), Fisher length $\delta x$ ($\times$), Onicescu length $\mathcal{L}_2$ ($\boxdot$), and Shannon length $N$ ($\odot$) of the Hermite polynomial $H_n(x)$ as a function of $n$.}
\label{fig_hermite}
\end{center}
\end{figure}

\begin{figure}
\begin{center}
\includegraphics[width=10cm]{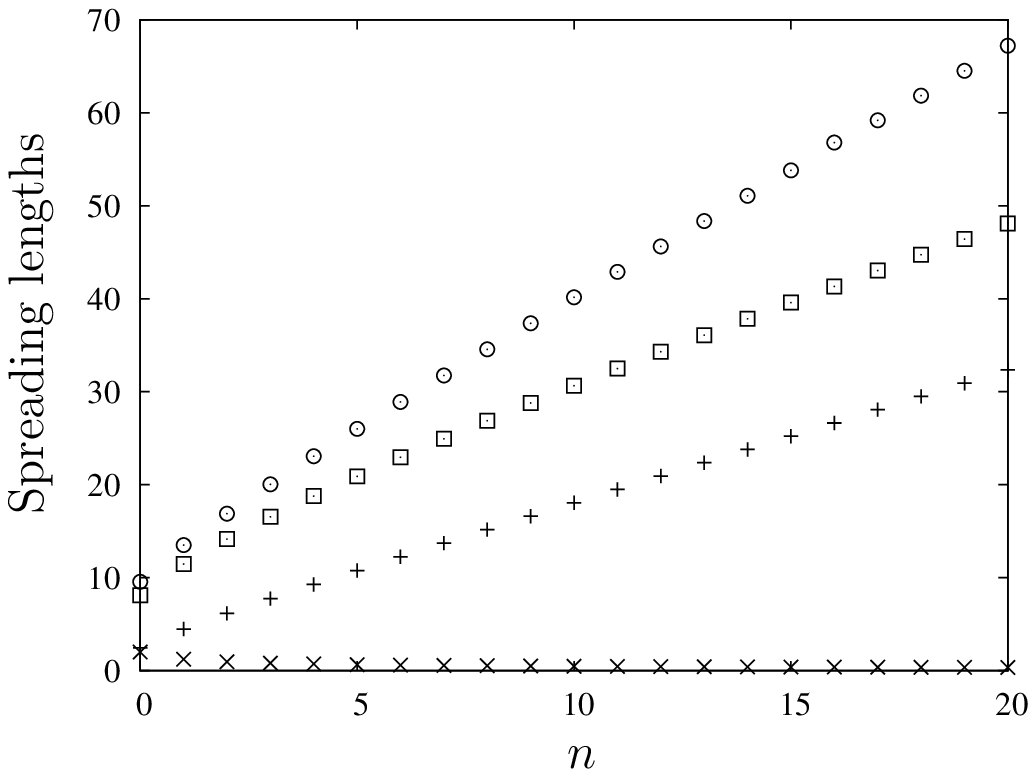}
\caption{Standard deviation $\Delta x$ ($+$), Fisher length $\delta x$ ($\times$), Onicescu length $\mathcal{L}_2$ ($\boxdot$), and Shannon length $N$ ($\odot$) of the Laguerre polynomial $L_n^{(5)}(x)$ as a function of $n$.}
\label{fig_laguerre}
\end{center}
\end{figure}

\begin{figure}
\begin{center}
\includegraphics[width=10cm]{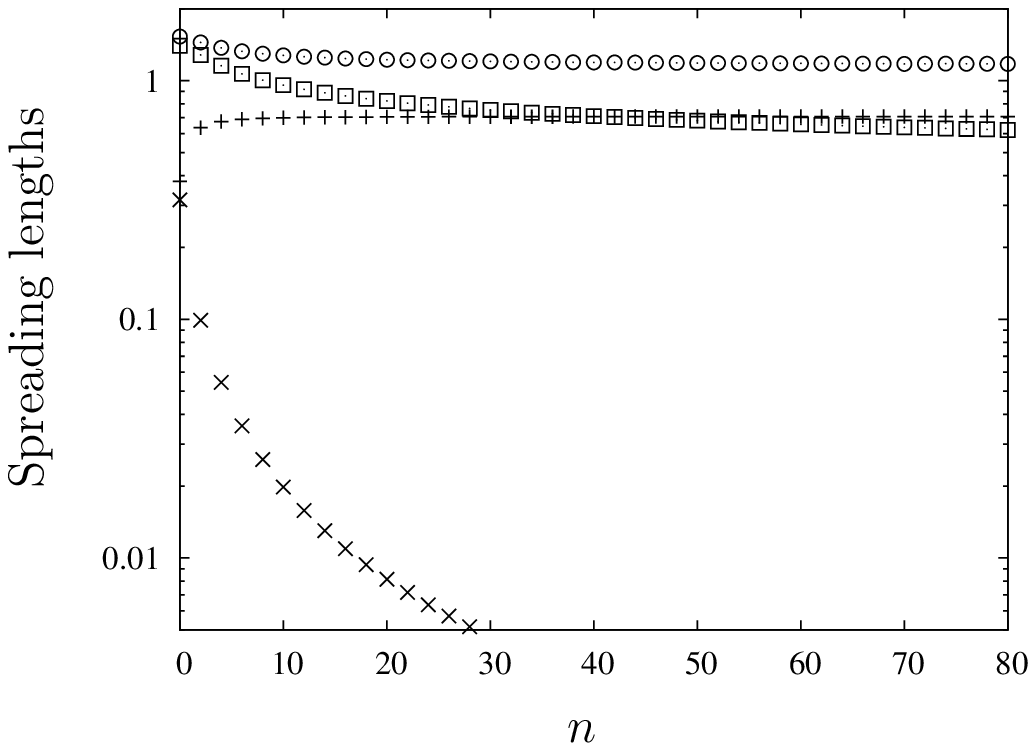}
\caption{Standard deviation $\Delta x$ ($+$), Fisher length $\delta x$ ($\times$), Onicescu length $\mathcal{L}_2$ ($\boxdot$), and Shannon length $N$ ($\odot$) of the Jacobi polynomial $P_n^{(2,2)}(x)$ as a function of $n$.}
\label{fig_jacobi}
\end{center}
\end{figure}

Here we carry out the numerical comparison of four direct spreading measures (standard deviation, Onicescu length or Renyi length with $q=2$, Shannon length and Fisher length) of various real classical orthogonal polynomials ($H_n(x)$, $L_n^{(5)}(x)$, $P_n^{(2,2)}(x)$) as a function of the degree $n$. This is done in Figures \ref{fig_hermite}, \ref{fig_laguerre} and \ref{fig_jacobi}.

In the Hermite and Laguerre cases all the four measures have a similar qualitative behaviour when the polynomial degree increases, as shown in Figures \ref{fig_hermite} and \ref{fig_laguerre}, respectively for $n=0,\ldots,20$. The three global quantities (standard deviation and Onicescu and Shannon lengths) grows when the degree $n$ is increasing, essentially because the polynomials spread more and more. Moreover, they behave so that $\Delta x < \mathcal{L}_2^R < N$, indicating that the spreading rate is weaker with respect to the mean value or centroid of the random variable than in absolute terms. In addition, we observe that the (local) Fisher length $\delta x$ decreases when the degree $n$ is increasing; this is because the polynomial becomes more and more oscillatory, so making its gradient content bigger. Finally, note that the Fisher length $\delta x<\Delta x$ in the two polynomial cases. It is worth remarking that the rate $\delta x/\Delta x$ fulfils the following characteristics. First, it is less than unity in accordance with the Cramer-Rao inequality (\ref{eq_cramer_rao_inequality}). Second, it decreases as $n^{-1}$ in the Hermite and as $n^{-\frac32}$ in the Laguerre case.

In the Jacobi case $P_n^{(2,2)}(x)$ the results are shown in Figure \ref{fig_jacobi} when $n$ goes from 0 to 80. The overall behaviour is qualitatively similar to the Hermite and Laguerre cases for the standard deviation and the Fisher length, while the Shannon and Onicescu lengths decrease in this case as $n$ grows. There are quantitative differences, of course; but there is one quantitative similarity: the rate $\delta x / \Delta x$ decreases as $n^{-\frac32}$ like in the Laguerre case. For a more extensive numerical discussion of the various spreading measures of these and other classical orthogonal polynomials, please see Refs. \cite{sanchezmoreno_jcam10}, \cite{sanchezmoreno_jcam11} and \cite{guerrero_jpa10}.

\section{Conclusions and open problems}

We have reviewed the knowledge of the recently introduced information-theoretic lengths of the classical orthogonal polynomials in a real, continuous variable. These measures quantify the spreading of these polynomials over its orthogonality interval in a complementary and more appropriate way than the standard deviation. They do not depend on any specific point of the interval. Moreover, they are direct spreading measures in the sense that they have the same units as the involved random variable and they possess a number of common invariance properties with the standard deviation, what facilitates their use and interpretation.

The standard deviation and the Fisher length can be explicitly given in a simple manner. The R\'enyi length can be expressed in terms of the degree $n$ and the polynomial parameter(s) via the Bell polynomials in all cases and a Lauricella function in the Laguerre case. The Shannon length has not yet been calculated but its asymptotics has been determined and some bounds have been found either variationally (Jacobi) or by means of an information-theoretic technique. Then, the accuracy of these bounds and the mutual comparision of the four direct spreading measures used in this work have been numerically studied. For completeness let us here mention that interesting but formal and cumbersome expressions between the Shannon entropy and the zeros of some classical orthogonal
polynomials have been found in the literature \cite{dehesa_jcam01,bata_1,sanchez_1}

Finally let us point out a few open problems whose solution is not only very relevant per se for the algebraic theory of classical orthogonal polynomials but they could also help to improve and extend this work. First, to derive linearization formulas for arbitrary powers of Hermite and Jacobi polynomials; this would help to find alternative expressions to the R\'enyi lengths of those polynomials. Second, to determine the $\mathcal{L}_q$-norms of the Laguerre and Jacobi polynomials what are not known for all the classical orthogonal polynomials save for the Hermite polynomials \cite{larsson:am02}.  These two problems and their corresponding asymptotics with respect to the degree of the polynomials and the parameter $q$ appear in the study of the R\'enyi lengths of the polynomials. Third, to find sharp variational bounds to the Shannon length of the orthogonal polynomials mentioned above. Fourth, to identify the largest class of orthogonal polynomials which satisfy the relation (\ref{eq:missing_one}) fulfilled by three canonical families of classical orthogonal polynomials in a real continuous variable. Fifth, to look for similar correlations among the Fisher and R\'enyi lengths and standard deviation for Hermite, Laguerre and Jacobi polynomials.

\subsection*{Acknowledgment}

This work has been partially funded by the Junta-de-Andaluc\'{\i}a grants FQM-207, FQM-2445 and FQM-4643 as well as the MICINN grant FIS2008-02380.

\bibliographystyle{unsrt}
\bibliography{krakow_paper}

\end{document}